# Interface Dipole : Effects on Threshold Voltage and Mobility for both Amorphous and Poly-crystalline Organic Field Effect Transistors.

C. Celle[1], C. Suspène[1], M. Ternisien[2*], S. Lenfant[2], D. Guérin[2], K. Smaali[2], K. Lmimouni[2], J.P. Simonato[1] and D. Vuillaume[2]

[1.] CEA,LITEN,DTNM,LCRE, 17 rue des Martyrs, F-38054 Grenoble, France
[2.] IEMN, CNRS & Univ. of Lille, UMR8520, Avenue Poincaré, Villeneuve d'Ascq, F-59652cedex, France.

E-mail : jean-pierre.simonato@cea.fr; dominique.vuillaume@iemn.univ-lille1.fr
Corresponding author :
D. Vuillaume (+33 320 19 78 66, dominique.vuillaume@iemn.univ-lille1.fr)

**Abstract.**

We report a detailed comparison on the role of a self-assembled monolayer (SAM) of dipolar molecules on the threshold voltage and charge carrier mobility of organic field-effect transistor (OFET) made of both amorphous and polycrystalline organic semiconductors. We show that the same relationship between the threshold voltage and the dipole-induced charges in the SAM holds when both types of devices are fabricated on strictly identical base substrates. Charge carrier mobilities, almost constant for amorphous OFET, are not affected by the dipole in the SAMs, while for polycrystalline OFET (pentacene) the large variation of charge carrier mobilities is related to change in the organic film structure (mostly grain size).

---

[*] Present/permanent address : Université de Toulouse; UPS, INPT, LAPLACE, 118 route de Narbonne, F-31062 Toulouse cedex 9, France.

# 1. Introduction

The control of the threshold voltage ($V_T$) of organic field-effect transistors (OFET) is still a key problem. Many groups have reported that the intercalation of a self-assembled monolayer (SAM) of molecules bearing a dipole between the gate dielectric and the organic semiconducting (OSC) film is an efficient way to modulate $V_T$ over a large range of values (more than few tens of Volt).[1-11] However, other OFET parameters (charge carrier mobility, on/off current ratio) can also be impacted by the SAMs due to their effects on (i) changes in the molecular orientation in the OSC, (ii) neutralization of surface defects, (iii) modification of surface roughness, interface dipole and surface energy. In most cases, a combination of these effects is likely, some of them being certainly dependent on both the dielectrics and the OSCs. As a consequence, several transistors parameters are modified simultaneously, and this feature prevents a clear understanding of the effect of the SAMs. We have previously demonstrated[9] that, in the case of an amorphous OSC, we can tune the threshold voltage alone, while keeping nearly unchanged the other electrical properties (hole carrier mobility, on/off ratio, subthreshold swing). Moreover, it is also difficult to precisely and quantitatively compare the results reported by various groups since, gate dielectric materials, OFET geometries, source and drain technologies are not systematically similar. For example, while a significant effect of SAMs on $V_T$ has been reported in the literature, experimental results on the effect of the SAMs on the charge carrier mobility are sparse or not discussed with respect to the dipole moment of molecules.[6, 8, 9]

Here, we report on a detailed comparison on the effects of dipolar SAMs on $V_T$ and charge carrier mobility for both amorphous (polytriarylamine, PTAA) and polycrystalline (pentacene, P5) OFET made on the same SAMs and the same transistor base substrates (same gate dielectric and source-drain electrode technology and geometry). We show that the same $V_T$ shifts are induced by the SAMs (10 different molecules with dipole moments ranging from -2D to 7D) for both PTAA and P5 OFETs. In particular, we observe that a linear correlation is obtained between $V_T$ and the dipole-induced charge in the SAM ($Q_{SAM}$) instead of the net dipole moment of the molecule. We also observe that the dipole moments of the molecules used in the SAMs have no significant effect on the charge carrier mobility.

## 2. Materials and Methods.

*2.1 Synthesis, monolayers, device fabrication.*

We used 10 molecules for SAM fabrication on the $SiO_2$ gate dielectric (Fig. 1). These molecules have dipole moments (along their long axis) ranging between ca. -2 D and 7 D (see Table 1, and calculation details below).

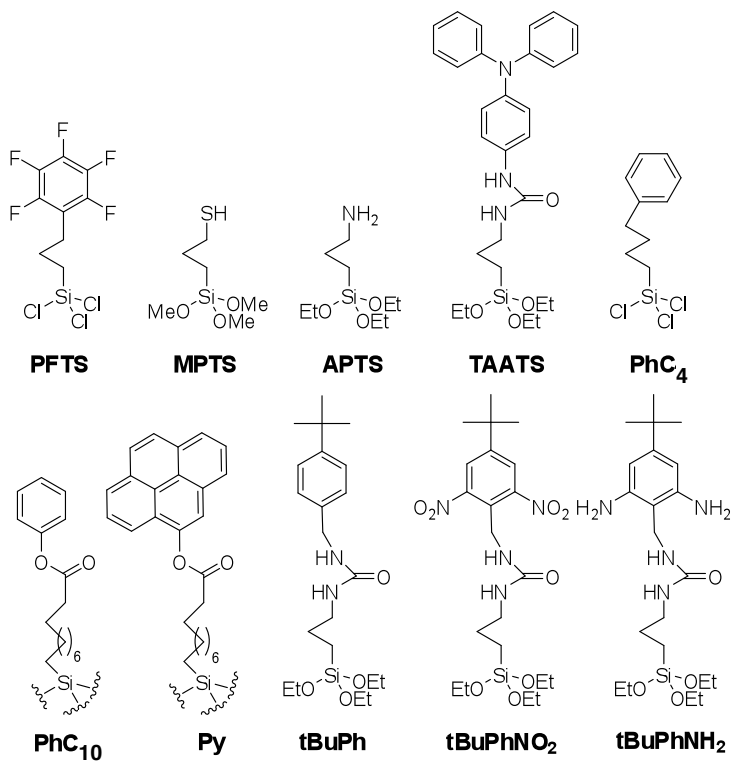

*Fig. 1* : *Chemical structures of dipolar molecules used in SAMs.*

**Molecules.**

**PFTS** ( $C_6F_5$-$(CH_2)_3$-$SiCl_3$ ; 3-pentafluorophenyl-propyl-trichloroSilane), **MPTS** (HS-$(CH_2)_3$-Si(OMe)$_3$ ; 3-mercaptopropyl-trimethoxysilane), **APTS** ($H_2N$-$(CH_2)_3$-Si(OEt)$_3$ ; 3-aminopropyl-triethoxysilane) and **PhC$_4$** ($C_6H_5$-$(CH_2)_4$-$SiCl_3$ ; 4-phenyl-butyl-trichlorosilane) were supplied by ABCR and used as received. **TAATS** (4-$(Ph_2N)$-$C_6H_4$-NHC(O)NH-$(CH_2)_3$-Si(OEt)$_3$ ; (1-(4-(diphenylamino)-phenyl)-3-(2-(triethoxysilyl)-propyl)-urea). This triarylamine derivative was synthesized in three steps according to a previously reported procedure.[9] **PhC$_{10}$** and **Py** were obtained through multistep synthesis directly on the surface according to a published protocol.[12] **tBuPh** (1-(4-tert-butylbenzyl)-3-(3-(triethoxysilyl)propyl)urea), **tBuPhNO$_2$** (1-(4-tert-butyl-2,6-dinitrobenzyl)-3-(3-(triethoxysilyl)propyl)urea) and **tBuPhNH$_2$** (1-(2,6-diamino-4-tert-butylbenzyl)-3-(3-(triethoxysilyl)propyl)urea were obtained by multistep synthesis as described in the supporting information.

**Self-assembled monolayers.**

Synthesis and characterization of the SAMs prepared from PFTS, MPTS, APTS, TAATS, Py and PhC$_{10}$ were described in our previous publications.[9, 12] In brief, n$^+$-doped silicon wafers were freshly cleaned and oxidized to provide a dense array of silanol groups (≡Si-OH), which are the anchoring sites for the organosilane molecules. Substrates were first cleaned by sonication in acetone, isopropanol then dichloromethane for 5 min. Wafers were dried under nitrogen flow then they were dipped into a freshly prepared piranha solution ($H_2SO_4$-$H_2O_2$ 2:1 v/v) at 100 °C for 30 min, or submitted to an oxygen plasma treatment (20 mTorr, 10 sccm $O_2$, 10 W, 300 s). They were rinsed thoroughly with deionized water then were dried under nitrogen stream. *Caution: the piranha solution reacts violently with organic chemicals. Consequently, it should be handled with extreme care*. The silanization reactions were carried out at room temperature in a nitrogen glovebox (<1 ppm $H_2O$ and $O_2$). For tBuPh, tBuPhNH$_2$, and tBuPhNO$_2$ SAMs, the freshly cleaned oxidized silicon substrates were immersed in a $10^{-3}$ M solution of the corresponding organosilane in anhydrous toluene then samples were kept in the dark for 4 days. Concerning PhC$_4$ SAM, the freshly cleaned silicon substrates were immersed for 2 h in a $10^{-3}$ M solution of PhC$_4$ in a mixture of hexane-dichloromethane 70:30 v/v. Functionalized substrates were cleaned in dichloromethane then isopropanol by sonication then blown with dry nitrogen.

**OFET**

The OFETs (both PTAA and P5) were processed using a bottom gate/bottom contact configuration. We used n$^+$-silicon (resistivity 1-3mΩ.cm) covered with a thermally grown 220 nm thick silicon dioxide (135 min at 1100°C in presence of oxygen 2L/min followed by a post-oxidation annealing at 900°C in $N_2$ 2L/min). The source and drain electrodes were patterned by optical lithography. Titanium/gold, (3/70 nm) were deposited by vacuum evaporation and lift-off. We fabricated linear and interdigitated OFETs with channel lengths L =1, 2, 5, 10, 20 and 50 µm and channel width W = 1,000 µm (linear OFET) and L= 2, 5, 10, 20 and 50 µm and W = 10,000 µm (interdigitated OFET). The back side of the silicon wafer was metalized by 200 nm thick aluminum to form an ohmic contact with the transistor gate. A 1wt% solution of PTAA (S1000 from Merck) in toluene was spin-coated on the substrates. Devices were annealed at 100°C for 20 min leading to 50 nm thick uniform thin films. Pentacene (P5 99%, purified 5 times, from Polysis) was vacuum sublimated at 0.01 nm/s with a nominal thickness of 30 nm).

*2.2. Film characterization.*

**Spectroscopic Ellipsometry:** We recorded spectroscopic ellipsometry data in the visible range using an UVISEL (Jobin Yvon Horiba) spectroscopic ellipsometer equipped with a DeltaPsi 2 data analysis software. The system acquired a spectrum ranging from 2 to 4.5 eV (corresponding to 300-750 nm) with intervals of 0.1 eV (or 15 nm). Data were taken at an angle of incidence of 70°, and the compensator was set at 45.0°. We fitted the data by a regression analysis to a film-on-substrate model as described by their thickness and their complex refractive indexes. Due to thickness variations of the thermally grown silicon dioxide, we used two samples during the SAM formation process: a thermally grown 200 nm thick silicon dioxide with the electrodes and a second silicon sample with a native oxide (1.0 to 1.5 nm thick). The SAM thickness was measured only on this second sample by spectroscopic ellipsometer. We assume that the SAM thicknesses measured on this second sample and on the sample with the thermally grown silicon dioxide are the same. To determine the monolayer thickness, we used the optical properties of silicon and silicon oxide from the software library, and for the monolayer we used the refractive index of 1.50. The usual values in the literature are in the range of 1.45–1.50.[13, 14] As a function of the wafer used, the native oxide layer thickness was measured between 10 and 15Å. The $SiO_2$ thickness was assumed to be unchanged after the monolayer assembly on the surface. We estimated the accuracy of the SAM thickness measurements at ± 1.5 Å. All measured thicknesses are in agreement with the calculated length of the molecules; meaning that we formed reasonably well-packed SAMs for all these 10 molecules.

**Contact Angle measurements:** We measured the static water contact angle with a remote-computer controlled goniometer system (DIGIDROP by GBX, France). We deposited a drop (10-30 μL) of deionized water (18 MΩ.cm$^{-1}$) on the surface, and the projected image was acquired and stored by the computer. Contact angles were extracted by contrast contour image analysis software. These angles were determined a few seconds after application of the drop. These measurements were carried out in a clean room (ISO 6) where the relative humidity (50%) and the temperature (22 °C) are controlled. The precision with these measurements are ±2°. All measured values (Table 1) are in agreement with literature data.[14, 15] For the 3 molecules ended by a tertiobutyl group, note that depending on the substituents of the aromatic ring, the hydrophobic behavior of monolayers decreased following the sequence: tBuPh > tBuPhNO$_2$ > tBuPhNH$_2$ (Table 1). One can observe that the buried polar substituents (NO$_2$ and NH$_2$) influence the surface energy of the monolayers despite the presence of the hydrophobic bulky tertiobutyl end group. Conformation of the molecules within the monolayers can explain this influence. Moreover steric hindrance of tertiobutyl probably decreases the compactness of the three SAMs.

**Kelvin Probe Force Microscopy**: For KPFM experiments we evaporated, on the SAM samples, clean platinum dots through a shadow mask (100 nm thick, with 100 x 100 µm dimensions), and samples were immediately transferred to the KPFM set-up and measured together. This protocol insures that the work function (WF) of the organic monolayer grafted on oxidized silicon is referenced to the same reference electrode (bare Pt dots). Since it is well known that metal atoms easily diffuse through organic monolayers and form many short-circuits, the WF of the upper Pt dots is thus a correct reference for the underlying silicon oxide substrate. KPFM measurements were carried out at room temperature with a Dimension 3100 placed under dry nitrogen atmosphere. We used Pt/Ir (0.95/0.05) metal-plated cantilevers with spring constant of ca. 3 N/m and a resonance frequency of ca. 70 kHz. Topography and KPFM data were recorded using a standard two-pass procedure, in which each topography line acquired in tapping mode is followed by the acquisition of KPFM data in a lift mode, with the tip scanned at a distance z ~ 80 nm above the sample so as to discard short range surface forces and be only sensitive to electrostatic forces. DC and AC biases ($V_{DC} + V_{AC} \sin(\omega t)$) are applied to the cantilever, and the platinum dot is electrically grounded ($V_{AC}$=2V, $\omega/2\pi$=70kHz). Experimentally, the contact potential difference (CPD) is measured using a feedback loop which sets to zero the cantilever oscillation amplitude at $\omega$ by adjusting the tip DC bias $V_{DC}$. This potential is simply equal to the CPD, in absence of side capacitances.[16]

**AFM:** The surface morphology of the organic layer was determined by imaging with a Veeco Dimension 3100 atomic force microscope (AFM) in tapping mode. Silicon cantilevers from MikroMasch (NSC15/ALBS) were used to acquire AFM images of 5 x 5 µm² at 1Hz with a resolution of 512 x 512 pixels.

**Dipole Calculation**: We calculated the dipole moment using MOPAC included in Chem3D 10.0 software (from Cambridge Soft Corporation, Cambridge, UK, 1996), with the semi-empirical method PM3. First, after an optimized conformation of the molecules (energy minimum optimization performed with MOPAC), the dipole values were calculated for isolated molecules with a $CH_3$ group at the end of the alkyl chain instead of the grafting part of the molecule (-Si-(OMe)$_3$, -Si-(OEt)$_3$, -Si-Cl$_3$). We considered that the leaving groups of the silane derivatives (OMe, OEt, Cl) are not present after the grafting reaction, and, therefore, no longer contribute to the molecular dipole. In the SAMs, the silyl groups form siloxane bonds, which are considered as part of the underline $SiO_2$ substrate. From these calculations on each individual molecule, the dipole values presented in table 1 are the projection of the dipole vector on the main axe of the molecule (this axe is defined by the alkyl part of the molecule in all-trans configuration).

*2.3. Transistor characterization.*

Electrical characterization of the OFETs were performed with an Agilent 4156B semiconductor parameter analyser. For the $I_D$-$V_G$ characteristics in the saturation regime, the back and forth gate voltage $V_G$ sweeps were carried out from 50V to -60V (40V to -40V for P5), with a step of 1V and at a drain bias of -60V (for PTAA) or -40V (for P5). Threshold voltage and charge carrier mobility were extrapolated, as usual, from a linear fit of the $I_D^{1/2}$ - $V_G$ plots. For each SAM, we fabricated several batches (max 3) of both PTAA and P5 OFETs and we systematically measured between 10 and 30 devices for each batch. In the following the reported $V_T$ and mobility values are the averaged values from all measured devices in each case.

## 3. Results and Discussion.

*3.1. Threshold voltage.*

OFETs made of PTAA and P5 with these 10 SAMs (see Fig. 1) were electrically characterized and the threshold voltage, $V_T$, extracted in the saturation regime. Fig. 2 shows typical drain current - gate voltage ($I_D$-$V_G$) and square-root of drain current vs. gate voltage ($I_D^{1/2}$ -$V_G$) curves for these devices.

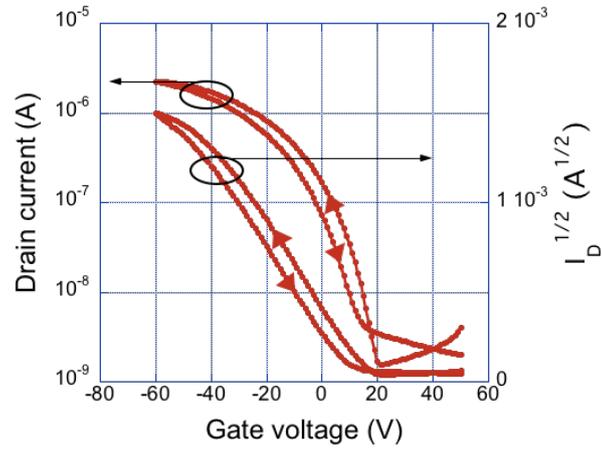

a

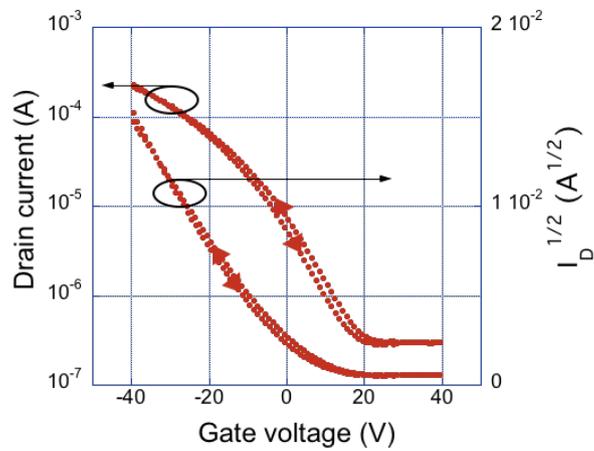

b

**Fig. 2 :** *Typical $I_D$-$V_G$ and $I_D^{1/2}$-$V_G$ characteristics for PTAA OFET (a) and P5 OFET (b) (L = 10 µm and W=1,000 µm in both cases) with a tBuPhNO$_2$ SAM.*

Following our previous approach[9] we plot in figure 3 the threshold voltage versus the dipole-induced charges in the SAM, $Q_{SAM}$, or more conveniently the surface density of charges $Q_{SAM}/e$, where e is the electron charge. For this set of data, $Q_{SAM}$ is calculated according to the Helmholtz equation which relates the potential drop, $V_{SAM}$, and the charge, $Q_{SAM}$, across a polar monolayer to the gas phase dipole moment, $P$, as:

$$V_{SAM} = \frac{NP_z}{\varepsilon_0 \varepsilon_{SAM}} \qquad (1)$$

and

$$Q_{SAM} = C_{SAM} V_{SAM} = \frac{NP_z}{t_{SAM}} \qquad (2)$$

where $N$ is the surface density of molecules in the SAM, $P_z$ is the dipole moment perpendicular to the substrate, $\varepsilon_0$ is the vacuum dielectric permittivity, $\varepsilon_{SAM}$ is the relative permittivity of the SAM, $t_{SAM}$ is the SAM thickness and $C_{SAM} = \varepsilon_0 \varepsilon_{SAM} / t_{SAM}$ is the capacitance of the SAM. As in our previous work,[9] we chose an average value of $2.5 \times 10^{14}$ cm$^{-2}$ ($\pm 1 \times 10^{14}$ cm$^{-2}$) for $N$, assuming a reasonable molecule packing in the SAMs (all other values for $P_z$ - calculated - and $t_{SAM}$ - ellipsometry measurements - are given in table 1). If we except the case of the tBuPhNO$_2$ SAM (which will be discussed later), the main conclusion here is that the linear relationship already observed for amorphous PTAA in Ref. [9] also holds for poly-crystalline P5 OFET made on the same Si/SiO$_2$ gate and Au source-drain structure. Solid lines in Fig. 3 are guide for eyes delimiting the lower and upper limits of the estimated correlation zone in this plot.

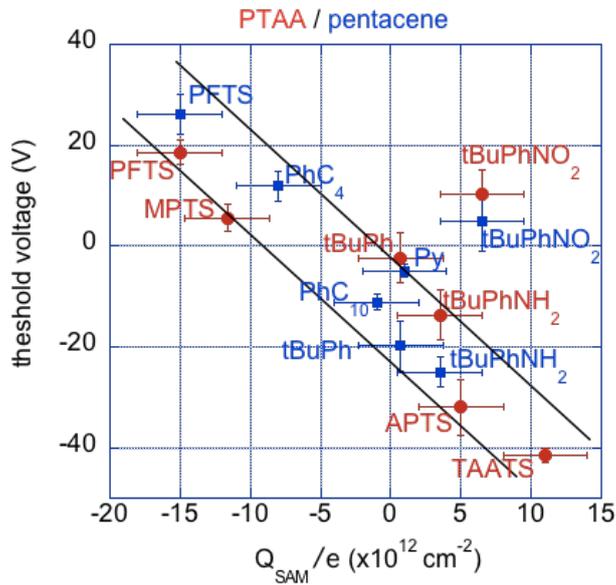

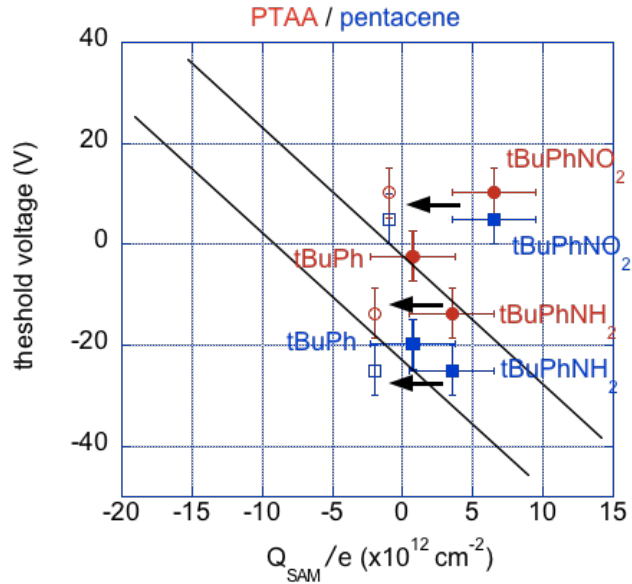

***Fig. 3 :*** *(a) Relationship between threshold voltage $V_T$ and dipole-induced surface density of charges in the SAM $Q_{SAM}/e$ (where e is the electron charge) for OFETs with the SAMs made of the 10 molecules shown in Fig. 1. Blue squares for P5 and red circles for PTAA. Solid lines are guide for eyes defining a correlation zone. (b) Comparison between data with $Q_{SAM}$ calculated from the molecular dipole (filled symbols) and measured from KPFM (unfilled symbols) for three SAMs : tBuPh, tBuPhNO$_2$ and tBuPhNH$_2$ (the data are superimposed for tBuPh, see table 1).*

Considering the hybrid gate dielectric (SAM on top of SiO$_2$, two capacitors in series), the relationship between the threshold voltage shift and the charge induced by the SAM can be written as

$$\Delta V_T = -\frac{C_{SAM} + C_{ox}}{C_{SAM} C_{ox}} Q_{SAM} \approx -\frac{Q_{SAM}}{C_{ox}} \qquad (3)$$

where $C_{SAM} = \varepsilon_0 \varepsilon_{SAM} / t_{SAM}$, the capacitance of the SAM (ranging from 1.3-3.1x10$^{-6}$ F/cm$^2$ depending on the SAM thickness, see table 1, and assuming $\varepsilon_{SAM}$=2.5) is larger than $C_{ox}$ = 1.7x10$^{-8}$ F/cm$^2$ (for the typical oxide gate dielectric of 220 nm). Thus, we get $\Delta V_T \approx$ -9.4x10$^{-12}$ $Q_{SAM}$. From Fig. 3-a, the slope (solid line) is about 2.7x10$^{-12}$ V.cm$^2$, a value smaller by a factor about 3.5. Many reasons can explain this difference. i) It is possible that the values of $Q_{SAM}$ are overestimated. The value of the molecular density, N in Eq. (2), is not exactly known in our case, and may be lowered if we have more disordered SAM. ii) The calculated values of $P_Z$ (table 1) are obviously overestimated. We do not take into account the depolarization effect, which is well known to significantly reduce the molecular dipole in a SAM (by a factor 3 -10)[17-19] compared to the value calculated for a single molecule in vacuum (the greater the dipole, the greater the depolarization effect). iii) In Eq. 3, we used an average value for $\varepsilon_{SAM}$ (=2.5), which is again a simplification since the exact value depends on the dipole moment of a given molecule. All these approximations lead to large error bars for $Q_{SAM}$ in Fig. 3.

The OFETs with the tBuPhNO$_2$ SAMs showed a significant deviation from the $V_T$-$Q_{SAM}$ correlation, $V_T$ being more positive than expected from the dipole-based calculated $Q_{SAM}$. This feature can be attributed to fixed negative charge trapped at the SAM/OSC interface[10] since NO$_2$ is a good electron acceptor group.

*Table 1: Main parameters of the SAMs used in this work. Individual molecule dipole ($P_z$) from MOPAC calculations, molecule length from geometry optimization, SAM thickness measured by ellipsometry, water contact angle, dipole-induced charges in the SAMs calculated from the molecule dipole and from KPFM experiments.*

| SAM | molecule dipole $P_z$ [1] (D) | molecule length [2] (nm) | SAM thickess [3] (nm) | Water CA [4] (°) | $Q_{SAM}/e$ from $P_z$ (cm$^{-2}$) [5] | $Q_{SAM}/e$ by KPFM (cm$^{-2}$) |
|---|---|---|---|---|---|---|
| PFTS | -1.80 | 1.00 | 0.90 | 92 | -1.5x10$^{13}$ | n.m. |
| MPTS | -1.60 | 0.77 | 0.75 | 65 | -1.2x10$^{13}$ | n.m. |
| APTS | 0.70 [6] | 0.74 | 0.71 | 72 | 5.0x10$^{12}$ | n.m. |
| TAATS | 7.70 [7] | 1.83 | n.m. | 89 | 1.1x10$^{13}$ | n.m. |
| Py | 0.15 | 1.85 | 1.80 | 74 | 2.0x10$^{11}$ | n.m. |
| PhC$_{10}$ | -0.10 | 1.90 | 1.80 | 75 | -3.0x10$^{11}$ | n.m. |
| PhC$_4$ | -0.30 | 1.20 | 1.10 | 85 | -8.2x10$^{12}$ | n.m. |
| tBuPh | 0.23 | 1.72 | 1.68 | 78 | 7.0x10$^{11}$ | 7.6x10$^{11}$ |
| tBuPhNO$_2$ | 2.70 | 1.73 | 1.60 | 70 | 8.2x10$^{12}$ | -2.0x10$^{12}$ |
| tBuPhNH$_2$ | 1.60 | 1.53 | 1.49 | 58 | 5.3x10$^{12}$ | -1.0x10$^{12}$ |

1. PM3 calculations
2. From geometry optimization PM3
3. Ellipsometry measurements (±0.15 nm)
4. ± 2°
5. Calculated with $\varepsilon_{SAM}$=2.5
6. protonated at 10% (see our previous paper[9])
7. cationic at 50% (see our previous paper[9])
n.m. : not measured

## 3.2. Monolayer voltage.

For some SAMs (tBuPh, tBuPhNO$_2$ and tBuPhNH$_2$), to obtain a better estimate of $Q_{SAM}$, we measured the monolayer voltage $V_{SAM}$ by Kelvin Probe Force Microscopy (KPFM). Fig. 4 shows a typical KPFM image and the KPFM histograms (see section 2 Materials and Methods for details) for these 3 SAMs compared to the bare SiO$_2$ surface, serving as reference. When the tip is over the Pt dots, the measured CPD is around zero (the Pt/Ir tip and Pt dots have almost the same WF). When the tip moves over the organic monolayer, we clearly observe a significant increase of the measured CPD, corresponding to a reduction of the WF of the functionalized substrate. Results in Fig. 4 clearly indicate that the WF of functionalized silicon oxidized is reduced by ca. 510mV, 340mV and 220mV for the tBuPh, tBuPhNH$_2$ and tBuPhNO$_2$ SAMs, respectively, compared to the bare Pt (we calculate the WF change from the maximum of the histogram peak). The corresponding $V_{SAM}$ values (the difference between CPD with the SAMs and the CPD of SiO$_2$) are 85 mV (tBuPh), - 230 mV (tBuPhNO$_2$) and - 115 mV (tBuPhNH$_2$). The related $Q_{SAM}$ values (calculated from Eq. 2) are given in Table 1. Except for the tBuPh SAM (for which $Q_{SAM}$ evaluated from the calculated dipole and by KPFM are almost the same, see Table 1), we note a shift of $Q_{SAM}$ measured by *KPFM* towards negative values (with respect to the calculated ones) for the tBuPhNO$_2$ and tBuPhNH$_2$ SAMs. This shift may be explained by one or more of the previously discussed features (dipole overestimation, trapped charges at the OSC/SAM interface). The data for these two SAMs are re-plotted as open symbols in Fig. 3-b. The general linear $V_T$ - $Q_{SAM}$ relationship remains valid, these values being still in the defined correlation zone. Moreover, in the case of the tBuPhNO$_2$ SAM, the value of Q$_{SAM}$ deduced from the KPFM experiments seems more relevant than the calculated one.

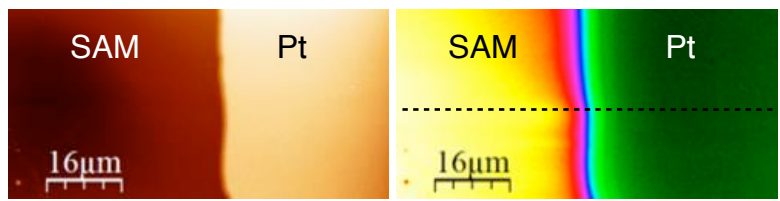

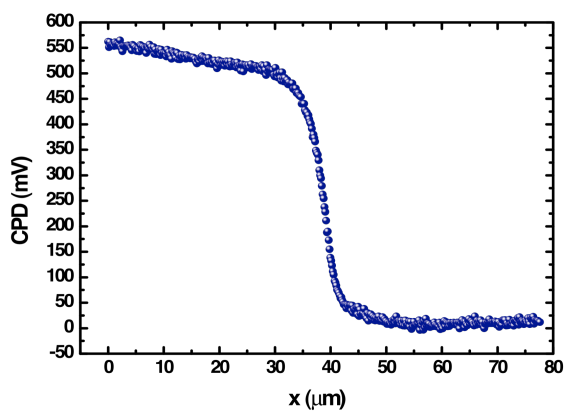

a

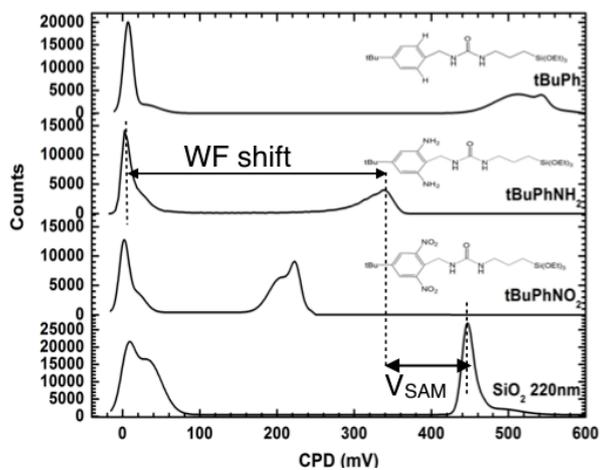

b

***Fig. 4.*** *(a) AFM (top left) and KPFM top right) images of the SAM/Pt area (tBuPh SAM) and corresponding KPFM profile along the dot line. (b) CPD histograms from KPFM images for 3 SAMs (tBuPh, tBuPhNO$_2$ and tBuPhNH$_2$) and bare SiO$_2$ (as reference for measuring V$_{SAM}$).*

Note that this KPFM evaluation on "naked" SAMs in air (no OSC on the top) are not accurately representative of the "full" device since charge trapping by the SAMs at the SAM/OSC interface is likely contributing to the threshold voltage shift as shown by Gholamrezaie at al.[10] Moreover, all our devices showed hysteresis in their $I_D$-$V_G$ curves of the order of 5 to15 V (Fig. 2). This hysteresis behavior can be due to trapping and detrapping of charges at the SAM/OSC interface with a dynamic on the same time scale as the voltage sweep. The observed hysteresis voltage value corresponds to charge trapping, $Q_T = V_H C_{ox}$ (with $V_H$ the amplitude of the *I-V* hysteresis), in the range 0.5-1.5 x $10^{12}$ cm$^{-2}$. This value is small compared to the large error on $Q_{SAM}$ (as explained above) and is neglected here (the $V_T$ values for the backward and forward traces of the I-V curves were not distinguished in the data analysis and averaged together). Note that the *I-V* hysteresis on the reference sample (SiO$_2$/OSC interface, no SAM) is generally smaller (< 5 V), thus a large part of these trapping arises at the SAM/OSC interface as shown in ref. [10].

*3.3. Charge carrier mobility.*

Charge carrier mobilities extracted in the linear regime are plotted versus $Q_{SAM}$ in Fig. 5.

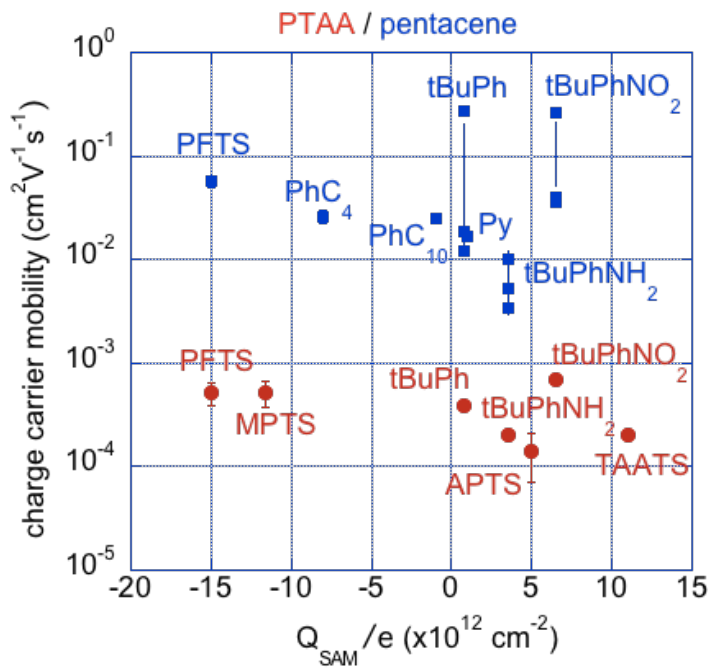

**Fig. 5**. *Charge carrier mobility for PTAA (red circles) and P5 (blue square) OFETs versus the dipole-induced surface density of charges in the SAMs $Q_{SAM}/e$ (where e is the electron charge).*

For PTAA OFET, we do not observe any significant dependance of the charge carrier mobility with the dipole of the molecules in the SAM. Data are almost constant within a factor ~ 5, a reasonable dispersion for organic transistors. For P5 OFET, we observe a larger dispersion (a factor ~ 100), with depends on both the nature of the SAM, but also within different batches of devices for the same SAM (see data for tBuPh, tBuPhNO$_2$ and tBuPhNH$_2$, linked by a thin vertical line). The 10 SAMs used in this work have different surface tension (see water contact angle in Table 1), and it is known that this parameter is one of the parameters that can affect the organization/order in the P5 film. We measured by AFM the size of grain domains of the P5 film between source and drain for the OFET with these different SAMs. Figure 6 shows typical AFM images of P5 films evaporated on tBuPh, tBuPhNO$_2$ and tBuPhNH$_2$ SAMs and on directly SiO$_2$. Data for other SAMs were already published.[20] The P5 average grain sizes were determined by 2D-FFT analysis using the WsXM software.[21] Fig. 7 shows a correlation plot of the mobility vs. the average P5 grain size.

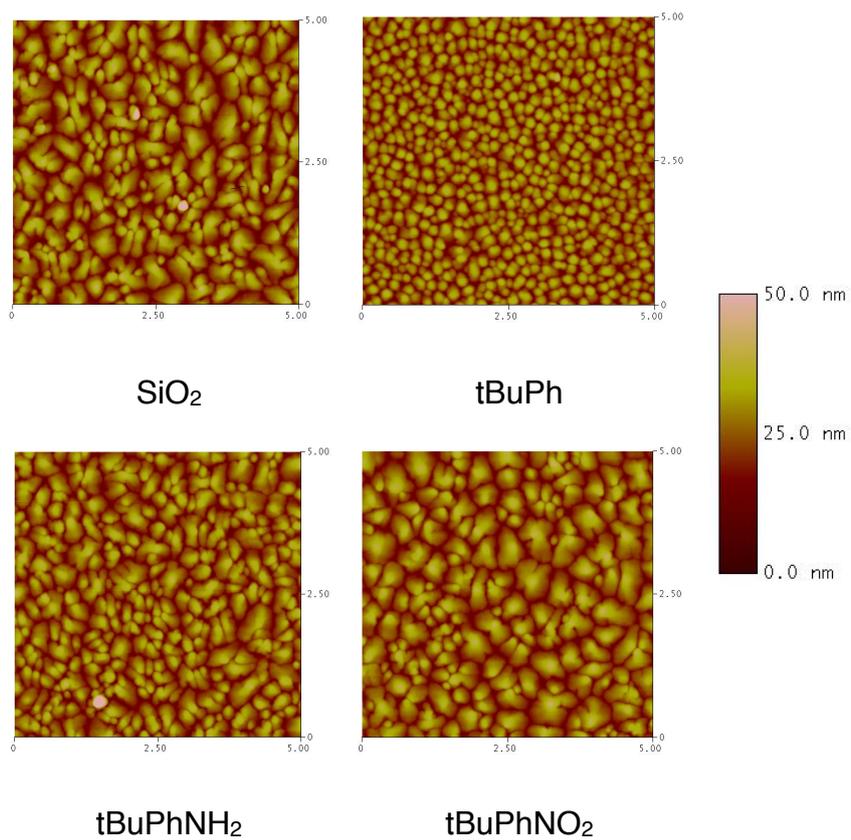

*Fig. 6*. *TM-AFM images of P5 films deposited on SAMs and bare SiO$_2$*

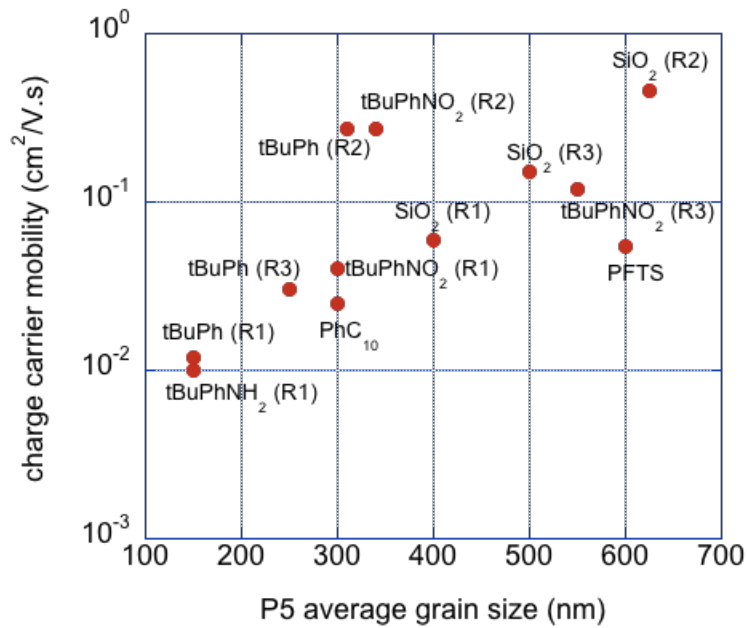

***Fig. 7.*** *Correlation between the charge carrier mobility in P5 OFET with various SAMs (and naked $SiO_2$ as reference) versus the average grain size measured by AFM as shown in Fig. 6.*

While we note a large dispersion in the mobility data and P5 grain size for several P5 deposition batches (labeled R1, R2 and R3) for a given SAM of the same chemical nature, we observe a general trend from this plot. As expected, the grain size is the main factor governing the charge carrier mobility inP5 OFET. Thus, we can conclude that the dipole moment of the SAMs has no significant influence on the charge carrier mobility, or is hidden by other factors, in particular here the correlation with changes in the structure properties of the organic film (grain size for P5). For the PTAA film, with an amorphous structure, mobility is roughly constant whatever the nature of the SAMs.

## 4. Conclusion.

In conclusion, we have shown that the threshold voltage of both amorphous and poly-crystalline OFETs are similarly controlled by SAMs of dipolar molecules (same relationship), while the charge carrier mobility is not. This relationship is broken in case of a strong charge trapping at the SAM/OSC interface, as exemplified by SAMs of molecules bearing a good electron acceptor groups. Charge carrier mobility is not affected by the dipole in the SAMs, remaining almost constant (within a factor 5) for amorphous OFET,, while for polycrystalline OFET (pentacene) the large variation of charge carrier mobilities (factor about 100) is related to change in the organic film structure (mostly grain size).

## Acknowledgements

Part of this work has been financially supported by the ministry of research (PhD grant of M.T.), the Agence National de la Recherche (CADISCOM project and post-doctoral grant for K.S.). The post-doctoral grant for C.S. was supported by the Chimtronique program of CEA.

# Reference

1. Kobayashi, S.; Nishikawa, T.; Takenobu, T.; Mori, S.; Shimoda, T.; Mitani, T.; Shimotani, H.; Yoshimoto, N.; Ogawa, S.; Iwasa, Y., Control of carrier density by self-assembled monolayers in organic field-effect transistors. *Nature Materials* **2004,** *3*, 317-322.

2. Chua, L.-L.; Ho, P. K. H.; Sirringhaus, H.; Friend, R. H., Observation of Field-Effect Transistor Behavior at Self-Organized Interfaces. *Adv. Mater.* **2004,** *16* (18), 1609-1615.

3. Chung, Y.; Verploegen, E.; Vailionis, A.; Sun, Y.; Nishi, Y.; Murmann, B.; Bao, Z., Controlling electric dipoles in nanodielectrics and its applications for enabling air-stable n-channel organic transistors. *Nano Letters* **2011,** *11* (3), 1161-1165.

4. Bruening, M.; Moons, E.; Yaron-Marcovitch, D.; Cahen, D.; Libman, J.; Shanzer, A., Polar ligand adsorption controls semiconductor surface potentials. *J. Am. Chem. Soc.* **1994,** *116* (7), 2972-2977.

5. Pernstich, K. P.; Haas, S.; Oberhoff, D.; Goldmann, C.; Gundlach, D. J.; Batlogg, B.; Rashid, A. N.; Schitter, G., Threshold voltage shift in organic field effect transistors by dipole monolayers on the gate insulator. *J. Appl. Phys.* **2004,** *96* (11), 6431.

6. Fleischli, F. D.; Suárez, S. p.; Schaer, M.; Zuppiroli, L., Organic Thin-Film Transistors: The Passivation of the Dielectric-Pentacene Interface by Dipolar Self-Assembled Monolayers. *Langmuir* **2010**, 100824135952023.

7. Ou-Yang, W.; Weis, M.; Taguchi, D.; Chen, X.; Manaka, T.; Iwamoto, M., Modeling of threshold voltage in pentacene organic field-effect transistors. *Journal of Applied Physics* **2010,** *107* (12), 124506.

8. Salinas, M.; Jäger, C. M.; Amin, A. Y.; Dral, P. O.; Meyer-Friedrichsen, T.; Hirsch, A.; Clark, T.; Halik, M., The Relationship between Threshold Voltage and Dipolar Character of Self-Assembled Monolayers in Organic Thin-Film Transistors. *Journal of the American Chemical Society* **2012**.

9. Celle, C.; Suspene, C.; Simonato, J. P.; Lenfant, S.; Ternisien, M.; Vuillaume, D., Self-assembled monolayers for electrode fabrication and efficient threshold voltage control of organic transistors with amorphous semiconductor layer. *Organic Electronics* **2009,** *10* (1), 119-126.

10. Gholamrezaie, F.; Andringa, A.-M.; Roelofs, W. S. C.; Neuhold, A.; Kemerink, M.; Blom, P. W. M.; De Leeuw, D. M., Charge Trapping by Self-Assembled Monolayers as the Origin of the Threshold Voltage Shift in Organic Field-Effect Transistors. *Small (Weinheim an der Bergstrasse, Germany)* **2011,** *8* (2), 241-245.

11. Ojima, T.; Koto, M.; Itoh, M.; Imamura, T., Control of field-effect transistor threshold voltages by insertion of self-assembled monolayers. *Journal of Applied Physics* **2013,** *113* (3), 034501.

12. Lenfant, S.; Guerin, D.; Tran Van, F.; Chevrot, C.; Palacin, S.; Bourgoin, J. P.; Bouloussa, O.; Rondelez, F.; Vuillaume, D., Electron transport through rectifying self-assembled monolayer diodes on silicon: Fermi-level pinning at the molecule-metal interface. *The Journal of Physical Chemistry B* **2006,** *110* (28), 13947-13958.

13. Parikh, A. N.; Allara, D. L.; Ben Azouz, I.; Rondelez, F., An intrinsic relationship between molecular structure in self-assembled n-alkylsiloxane monolayers and deposition temperature. *J. Phys. Chem.* **1994,** *98*, 7577-7590.

14. Ulman, A., *An introduction to ultrathin organic films : from Langmuir-Blodgett to Self-assembly*. Academic press: Boston, 1991.

15. Schreiber, F., Structure and growth of self-assembling monolayers. *Progress in Surf. Sci.* **2000,** *65* (5-8), 151-256.

16. Brunel, D.; Deresmes, D.; Mélin, T., Determination of the electrostatic lever arm of carbon nanotube field effect transistors using Kelvin force microscopy. *Applied Physics Letters* **2009,** *94* (22), -.

17. Krzeminski, C.; Delerue, C.; Allan, G.; Vuillaume, D.; Metzger, R., Theory of electrical rectification in a molecular monolayer. *Physical Review B* **2001,** *64* (8), 085405.

18. Natan, A.; Kronik, L.; Haick, H.; Tung, R. T., Electrostatic properties of ideal and non-ideal polar organic monolayers: implications for electronic devices. *Adv. Mat.* **2007**, *19*, 4103-4117.

19. Cornil, D.; Olivier, Y.; Geskin, V.; Cornil, J., Depolarization Effects in Self-Assembled Monolayers: A Quantum-Chemical Insight. *Adv. Funct. Mater.* **2007**, *17* (7), 1143-1148.

20. Ternisien, M. Contribution à l'étude des transistors à effet de champ organiques : Effet de la fonctionnalisation du diélectrique de grille par des monocouches auto-assemblées. PhD, University of Lille, 2008.

21. Horcas, I.; Fernández, R.; Gómez-Rodríguez, J. M.; Colchero, J.; Gómez-Herrero, J.; Baro, A. M., WSXM: A software for scanning probe microscopy and a tool for nanotechnology. *Rev. Sci. Instrum.* **2007**, *78* (1), 013705.

# Interface Dipole : Effects on Threshold Voltage and Mobility for both Amorphous and Poly-crystalline Organic Field Effect Transistors.

C. Celle (1), C. Suspène (1), S. Lenfant (2), D. Guérin (2), K. Smaali (2), M. Ternisien (2)[*], K. Lmimouni (2), J.P. Simonato (1) & D. Vuillaume (2)

1) CEA/LITEN/DTNM/LCRE, 17 rue des Martyrs, F-38054 Grenoble, France
2) IEMN, CNRS & Univ. of Lille, UMR8520, Avenue Poincaré, Villeneuve d'Ascq, F-59652 cedex, France.

E-mail : jean-pierre.simonato@cea.fr; dominique.vuillaume@iemn.univ-lille1.fr

## Supplementary Materials

**1- Synthesis of molecular precursors for the fabrication of SAMs**

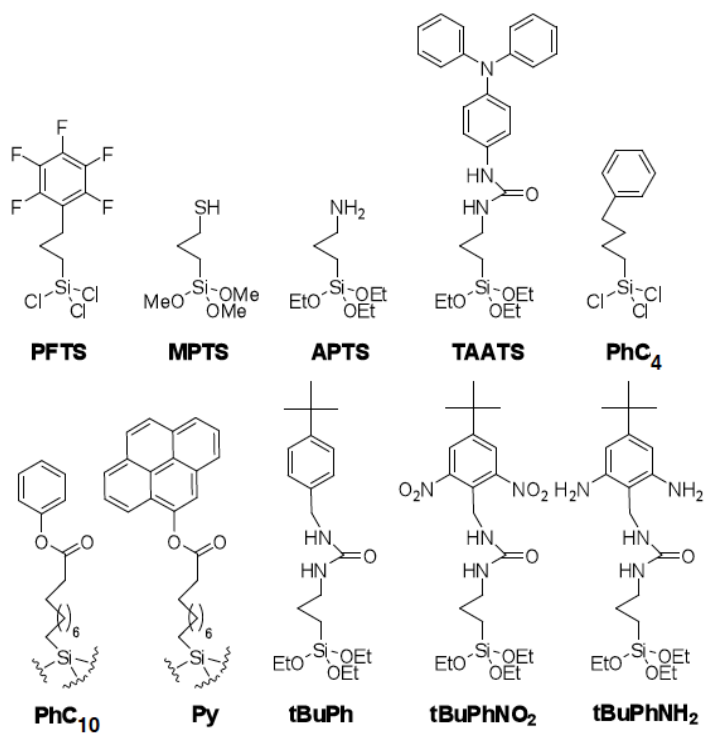

**Figure S1.** Chemical structures of various molecular precursors for the fabrication of SAMs.

---

**Synthesis of tBuPh.** 1-(4-tert-butylbenzyl)-3-(3-(triethoxysilyl)propyl)urea

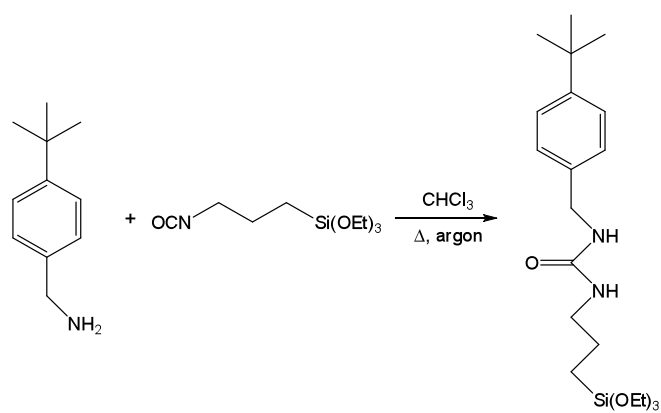

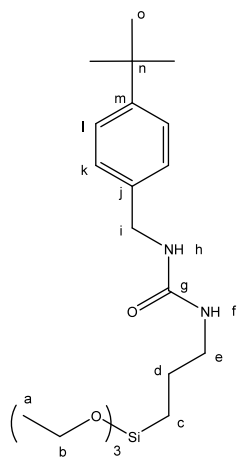

217 mL (1.23 mmol) of 4-*tert*-butylbenzylamine were diluted in 15 mL of chloroform in a 50 mL round-bottom flask equipped with a cooler and a magnetic stirring bar. The vessel was sealed with a rubber septum, evacuated and backfilled with argon. 267 µL (1.08 mmol) of (3-isocyanatopropyl)triethoxysilane were then rapidly added with a syringe and the mixture was heated with stirring at reflux under argon for 20 h. After cooling down to room temperature, the solvent was removed in vacuum to afford the desired compound as a pale yellow oil (440 mg, yield: 87 %). It was possible to purify the silane by flash chromatography on SiO$_2$ column using a mixture of hexane-AcOEt 50:50 v/v as eluent (white solid, 250 mg, yield: 50% after chromatography). tBuPh was stored under nitrogen.

$^1$H NMR (200 MHz, CDCl$_3$, 300 K): $\delta$ (ppm) = 0.65 (t, 2H, $^3J$ = 8.3 Hz, H$_c$), 1.24 (t, 9H, $^3J$ = 7.3 Hz, H$_a$), 1.35 (s, 9H, H$_o$), 1.65 (p, 2H, $^3J$ = 8.3 Hz, H$_d$), 3.18 (dt, 2H, $^3J$ = 8.3, 5.8 Hz, H$_e$), 3.85 (q, 6H, $^3J$ = 7.3 Hz, H$_b$), 4.32 (d, 2H, $^3J$ = 5.9 Hz, H$_i$), 4.79 (t, 1H, $^3J$ = 5.8 Hz, H$_f$), 4.92 (t, 1H, $^3J$ = 5.9 Hz, H$_h$), 7.32 (m, 4H, H$_k$ + H$_l$).

$^{13}$C{$^1$H} NMR (200 MHz, CDCl$_3$, 300 K): $\delta$ (ppm) = 7.8 (C$_c$), 18.2 (C$_a$), 23.7 (C$_d$), 31.2 (C$_o$), 34.5 (C$_n$), 42.8 (C$_e$), 44.1 (C$_i$), 58.3 (C$_b$), 125.5 (C$_l$), 127.1 (C$_k$), 136.3 (C$_j$), 150.2 (C$_m$), 158.5 (C$_g$).

IR (ATR, cm$^{-1}$): 3341 ($\nu$ N-H), 3062 ($\nu$ C-H$_{Ar}$), 2968 ($\nu_{as}$ C-H), 2927 ($\nu_s$ C-H), 1631 ($\nu$ C=O), 1571 (n C=C$_{Ar}$), 1514 ($\delta$ N-H), 1271 ($\nu$ C-N), 1235 ($\nu$ C-C), 1078 ($\nu_{as}$ Si-O-C), 956 ($\nu_s$ Si-O-C).

**Synthesis of tBuPhNO$_2$:** 1-(4-*tert*-butyl-2,6-dinitrobenzyl)-3-(3-(triethoxysilyl)propyl)urea

This synthesis is performed in 6 steps:

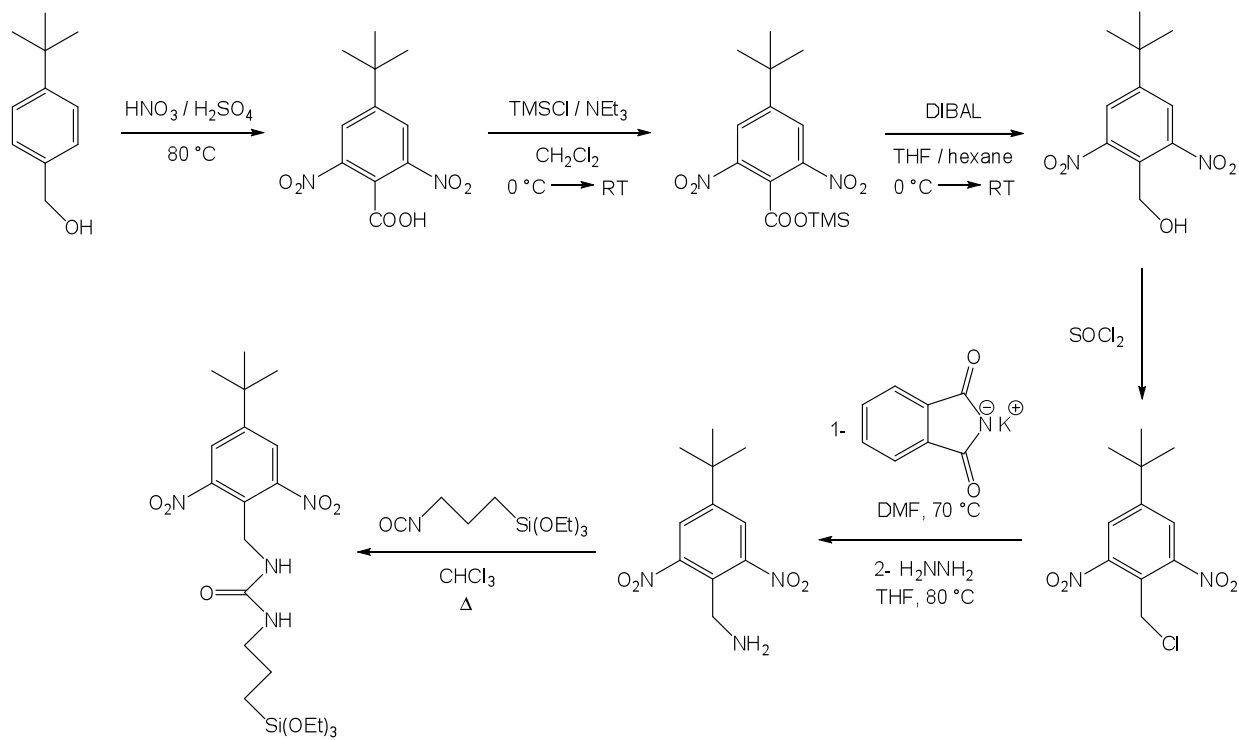

Each step is described in the following section.

**4-*tert*-butyl-2,6-dinitrobenzoic acid**

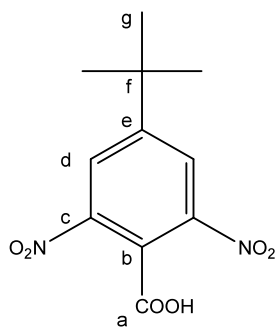

24 mL of sulfuric acid (100 %) were slowly added with stirring at 0 °C to 18 mL of nitric acid (100 %) placed in a 100 mL round-bottom flask. In parallel, 4.3 mL (24.4 mmol) of 4-*tert*-butylbenzyl alcohol were placed in another 100 mL round-bottom flask. The mixture of fuming acids was then slowly added to the alcohol at 0 °C. The resulting solution was stirred at room temperature for 2 h and at 80 °C for 3 h and was then poured into 200 mL of iced water. The precipitate was filtered, copiously rinsed with water and dried under vacuum to afford the desired compound as a yellow solid (5.2 g, yield: 77 %).

$^{1}$H NMR (200 MHz, CDCl$_3$, 300 K): $\delta$ (ppm) = 1.41 (s, 9H, H$_g$), 8.26 (s, 2H, H$_d$), 15.63 (s, 1H, H$_a$).

$^{13}$C{$^{1}$H} NMR (200 MHz, CDCl$_3$, 300 K): $\delta$ (ppm) = 29.1 (C$_g$), 37.2 (C$_f$), 127.2 (C$_d$), 131.8 (C$_b$), 137.8 (C$_c$), 152.1 (C$_e$), 163.8 (C$_a$).

IR (ATR, cm$^{-1}$): 3075 ($\nu$ C-H$_{Ar}$), 2972 ($\nu_{as}$ C-H), 2930 ($\nu_s$ C-H), 2650 ($\nu$ O-H), 1712 ($\nu$ C=O), 1588 (n C=C$_{Ar}$), 1542 ($\nu$ NO$_2$), 1360 ($\nu$ NO$_2$), 1288 ($\nu$ C-O).

**(4-*tert*-butyl-2,6-dinitrophenyl)methanol**

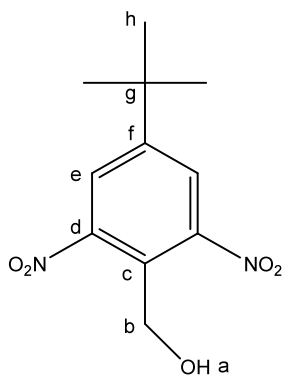

5.2 g (19.4 mmol) of 4-*tert*-butyl-2,6-dinitrobenzoic acid were dissolved in 20 mL of dichloromethane in a 500 mL round-bottom flask. Then, 2.7 mL (19.4 mmol) of triethylamine and 2.5 mL (19.4 mmol) of trimethylsilyl chloride were successively added with stirring at 0 °C. The mixture was allowed to warm to room temperature and was stirred for 3 h. In parallel, 58.6 mL (58.2 mmol) of a solution of diisobutylaluminium hydride (1 M in tetrahydrofuran) were added dropwise to 20 mL of tetrahydrofuran at 0 °C in another 500 mL round-bottom flask previously evacuated and backfilled with argon. The solution containing the trimethylsilyl-protected acid was then slowly added to the solution of diisobutylaluminium hydride in tetrahydrofuran at 0 °C. The resulting mixture was allowed to warm to room temperature and was stirred for 24 h under argon. Another equivalent (58.6 mL) of a solution of diisobutylaluminium hydride (1 M in tetrahydrofuran) was next introduced and the solution was stirred at room temperature for 48 h under argon. The reaction was quenched by addition of 50 mL of methanol and solvents were removed under vacuum. The residue was redissolved in 20 mL of methanol and solid impurities were filtered off. 200 mL of a saturated solution of ammonium chloride were then added and the organic phase was extracted three times with 25 mL of dichloromethane and filtered over Celite®. Drying over magnesium sulfate, filtration and evaporation of solvents afforded the desired compound as a red-brown solid (4.5 g, yield: 91 %).

$^1$H NMR (200 MHz, CDCl$_3$, 300 K): $\delta$ (ppm) = 1.48 (s, 9H, H$_h$), 2.02 (s$_{br}$, 1H, H$_a$), 4.72 (s, 2H, H$_b$), 7.52 (s, 2H, H$_e$).

$^{13}$C{$^1$H} NMR (200 MHz, CDCl$_3$, 300 K): $\delta$ (ppm) = 31.3 (C$_h$), 38.2 (C$_g$), 62.9 (C$_b$), 124.6 (C$_e$), 129.2 (C$_c$), 142.1 (C$_d$), 151.1 (C$_f$).

IR (ATR, cm$^{-1}$): 3369 ($\nu$ O-H), 3062 ($\nu$ C-H$_{Ar}$), 2978 ($\nu_{as}$ C-H), 2927 ($\nu_s$ C-H), 1592 (n C=C$_{Ar}$), 1538 ($\nu$ NO$_2$), 1364 ($\nu$ NO$_2$), 1071 ($\nu$ C-O).

**5-*tert*-butyl-2-(chloromethyl)-1,3-dinitrobenzene**

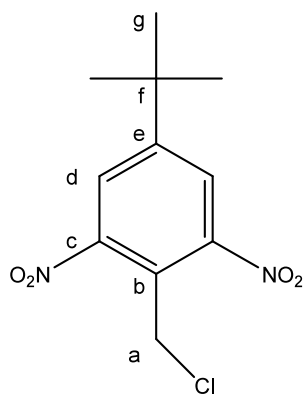

4.5 g (17.7 mmol) of (4-*tert*-butyl-2,6-dinitrophenyl)methanol were dissolved in 20 mL of dichloromethane in a 100 mL round-bottom flask. 1.6 mL (21.2 mmol) of thionyl chloride were then added and the mixture was stirred at room temperature for 4 h. The excess of thionyl chloride and the solvent were removed in vacuo to afford a red-brown solid (4.5 g, yield: 93 %).

$^1$H NMR (200 MHz, CDCl$_3$, 300 K): $\delta$ (ppm) = 1.49 (s, 9H, H$_g$), 4.54 (s, 2H, H$_a$), 7.53 (s, 2H, H$_d$).

$^{13}$C{$^1$H} NMR (200 MHz, CDCl$_3$, 300 K): $\delta$ (ppm) = 30.8 (C$_g$), 37.2 (C$_f$), 42.6 (C$_a$), 124.3 (C$_b$), 126.3 (C$_d$), 141.6 (C$_c$), 150.7 (C$_e$).

IR (ATR, cm$^{-1}$): 3065 ($\nu$ C-H$_{Ar}$), 2973 ($\nu_{as}$ C-H), 2925 ($\nu_s$ C-H), 1589 (n C=C$_{Ar}$), 1538 ($\nu$ NO$_2$), 1363 ($\nu$ NO$_2$).

**2-(4-*tert*-butyl-2,6-dinitrobenzyl)isoindoline-1,3-dione**

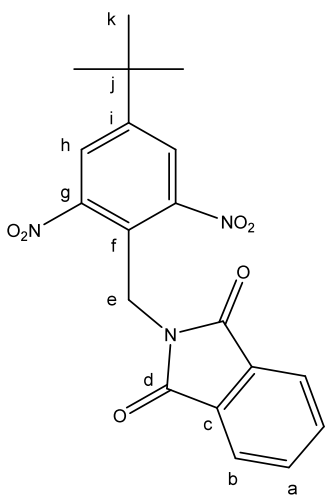

2.5 g (9.2 mmol) of 5-*tert*-butyl-2-(chloromethyl)-1,3-dinitrobenzene were dissolved in 20 mL of dimethylformamide in a 100 mL round-bottom flask. 1.9 g (10.2 mmol) of phthalimide potassium salt were then added and the mixture was stirred at 80 °C for 24 h. The resulting solution was poured into 150 mL of water and the organic phase was extracted three times with 20 mL of dichloromethane, intensively rinsed with water and dried over magnesium sulfate. Filtration and evaporation of the solvent afforded the desired compound as an orange solid (1.7 g, yield: 49 %)

$^1$H NMR (200 MHz, CDCl$_3$, 300 K): $\delta$ (ppm) = 1.46 (s, 9H, H$_k$), 4.84 (s, 2H, H$_e$), 7.54 (s, 2H, H$_h$), 7.75 and 7.92 (m, 4H, H$_a$ + H$_b$).

$^{13}$C{$^1$H} NMR (200 MHz, CDCl$_3$, 300 K): $\delta$ (ppm) = 31.3 (C$_k$), 38.7 (C$_j$), 43.1 (C$_e$), 124.3 (C$_f$), 126.4 (C$_h$), 127.3 (C$_b$), 131.4 (C$_c$), 133.4 (C$_a$), 141.7 (C$_g$), 150.9 (C$_i$), 169.1 (C$_d$).

IR (ATR, cm$^{-1}$): 3072 (ν C-H$_{Ar}$), 2975 (ν$_{as}$ C-H), 2926 (ν$_s$ C-H), 1774 and 1714 (ν C=O), 1590 (n C=C$_{Ar}$), 1539 (ν NO$_2$), 1368 (ν NO$_2$), 1294 (ν C-N).

**(4-*tert*-butyl-2,6-dinitrophenyl)methanamine**

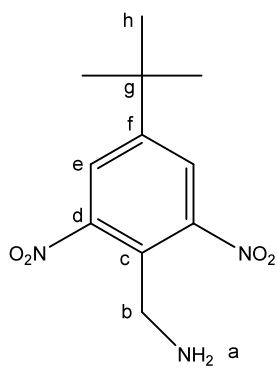

1.2 g (3.1 mmol) of 2-(4-*tert*-butyl-2,6-dinitrobenzyl)isoindoline-1,3-dione were dissolved in 20 mL of tetrahydrofuran in a 100 mL round-bottom flask. 761 mL (15.7 mmol) of hydrazine hydrate were then added and the mixture was stirred at 80 °C for 3 h. After cooling down to room temperature, 10 mL of a solution of hydrochloric acid (37 %) were introduced and the mixture was stirred at 80 °C for 1 h. The precipitate was then filtered and copiously rinsed with water. After deprotonation by addition of sodium hydroxide in 30 mL of an ethanol/water 1:1 mixture, the organic phase was extracted three times with 10 mL of dichloromethane and dried over magnesium sulfate. Filtration, evaporation of the solvent and recrystallization from acetone afforded pale yellow crystals (620 mg, yield: 78 %).

$^1$H NMR (200 MHz, CDCl$_3$, 300 K): $\delta$ (ppm) = 1.48 (s, 9H, H$_h$), 1.63 (s$_{br}$, 2H, H$_a$), 3.95 (s, 2H, H$_b$), 7.51 (s, 2H, H$_e$).

$^{13}$C{$^1$H} NMR (200 MHz, CDCl$_3$, 300 K): $\delta$ (ppm) = 31.3 (C$_h$), 38.2 (C$_g$), 45.8 (C$_b$), 124.9 (C$_e$), 129.5 (C$_c$), 142.1 (C$_d$), 151.2 (C$_f$).

IR (ATR, cm$^{-1}$): 3469 (ν N-H), 3375 (ν N-H), 3063 (ν C-H$_{Ar}$), 2973 (ν$_{as}$ C-H), 2924 (ν$_s$ C-H), 1588 (n C=C$_{Ar}$), 1537 (ν NO$_2$), 1497 (δ N-H), 1364 (ν NO$_2$), 1274 (ν C-N).

**Synthesis of tBuPhNO$_2$:** 1-(4-*tert*-butyl-2,6-dinitrobenzyl)-3-(3-(triethoxysilyl)propyl)urea

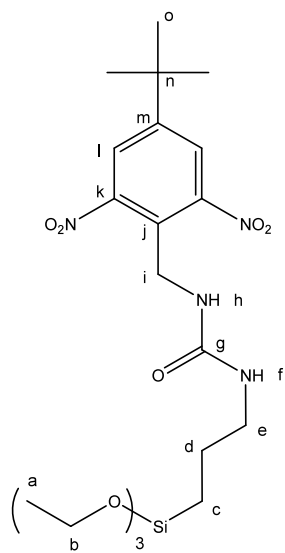

250 mg (0.99 mmol) of (4-*tert*-butyl-2,6-dinitrophenyl)methanamine were diluted in 10 mL of chloroform in a 50 mL round-bottom flask equipped with a cooler and a magnetic stirring bar. The vessel was sealed with a rubber septum, evacuated and backfilled with argon. 270 μL (1.08 mmol) of (3-isocyanatopropyl)triethoxysilane were then rapidly added with a syringe and the mixture was heated with stirring at reflux under argon for 20 h. After cooling down to room temperature, the solvent was removed in vacuo and the oily residue poured into 20 mL of pure pentane (> 99 %). After sonication and decantation, the supernatant was removed with a syringe and then the residue was dissolved in 5 mL of dichloromethane and transferred into a Schlenk tube with a cannula. The solvent was finally evaporated to afford the desired compound as a brown solid (300 mg, yield: 61 %). For the preparation of SAMs, the silane could be further purified by flash chromatography on $SiO_2$ using a mixture of $CHCl_3$-AcOEt 70:30 v/v as an eluent. A pale brown solid was isolated and stored under nitrogen.

$^1$H NMR (200 MHz, DMSO, 300 K): $\delta$ (ppm) = 0.51 (t, 2H, $^3J$ = 8.2 Hz, $H_c$), 1.16 (t, 9H, $^3J$ = 7.3 Hz, $H_a$), 1.37 (s, 9H, $H_o$), 1.61 (p, 2H, $^3J$ = 8.2 Hz, $H_d$), 2.95 (dt, 2H, $^3J$ = 8.2, 5.7 Hz, $H_e$), 3.71 (q, 6H, $^3J$ = 7.3 Hz, $H_b$), 4.22 (d, 2H, $^3J$ = 5.9 Hz, $H_i$), 6.16 (t, 1H, $^3J$ = 5.7 Hz, $H_f$), 6.51 (t, 1H, $^3J$ = 5.9 Hz, $H_h$), 7.67 (s, 2H, $H_l$).

$^{13}$C{$^1$H} NMR (200 MHz, DMSO, 300 K): $\delta$ (ppm) = 7.6 ($C_c$), 18.4 ($C_a$), 24.5 ($C_d$), 31.0 ($C_o$), 34.3 ($C_n$), 42.9 ($C_e$), 43.1 ($C_i$), 58.5 ($C_b$), 125.2 ($C_l$), 129.3 ($C_j$), 142.3 ($C_k$), 150.8 ($C_m$), 159.4 ($C_g$).

IR (ATR, cm$^{-1}$): 3356 (ν N-H), 3061 (ν C-H$_{Ar}$), 2972 (ν$_{as}$ C-H), 2927 (ν$_s$ C-H), 1638 (ν C=O), 1569 (n C=C$_{Ar}$), 1539 (ν $NO_2$), 1474 (δ N-H), 1365 (ν $NO_2$), 1251 (ν C-N), 1072 (ν$_{as}$ Si-O-C), 954 (ν$_s$ Si-O-C).

**Synthesis of tBuPhNH$_2$ :** 1-(2,6-diamino-4-*tert*-butylbenzyl)-3-(3-(triethoxysilyl)propyl)urea

This synthesis was realized in seven steps. The molecule (4-*tert*-butyl-2,6-dinitrophenyl)methanamine is an intermediate for the syntheses of both **tBuPhNO₂** and **tBuPhNH₂**.

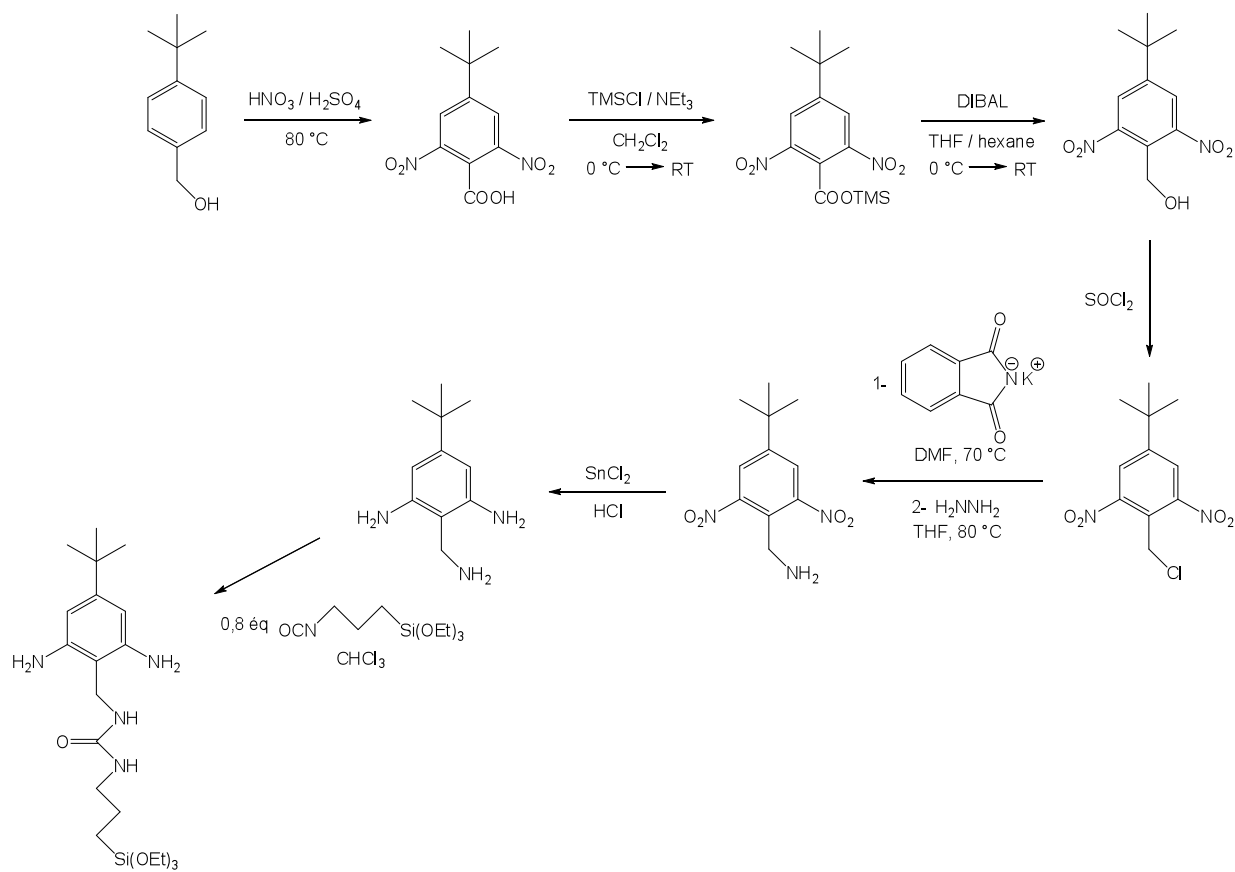

## 2-(aminomethyl)-5-*tert*-butylbenzene-1,3-diamine

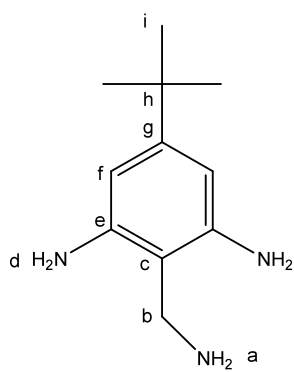

200 mg (0.79 mmol) of (4-*tert*-butyl-2,6-dinitrophenyl)methanamine were dissolved in 15 mL of a solution of hydrochloric acid (37 %) in a 50 mL round-bottom flask. 1.4 g (6.3 mmol) of tin(II) chloride dihydrate dissolved in 10 mL of a solution of hydrochloric acid (37 %) were then slowly added and the mixture was stirred at 70 °C for 15 h. After cooling down to room temperature and deprotonation by addition of sodium hydroxide until pH > 12, the organic phase was extracted three times with 10 mL of dichloromethane and dried over magnesium sulfate. Filtration, evaporation of the solvent and recrystallization from pentane afforded a pale yellow solid (97 mg, yield: 63 %). Caution: the product was sensitive to light and should be stored in the dark under nitrogen.

$^1$H NMR (200 MHz, DMSO, 300 K): $\delta$ (ppm) = 1.35 (s, 9H, H$_i$), 2.21 (s$_{br}$, 2H, H$_a$), 3.63 (s, 2H, H$_b$), 5.91 (s, 2H, H$_f$), 6.12 (s, 4H, H$_d$).

$^{13}$C{$^1$H} NMR (200 MHz, DMSO, 300 K): $\delta$ (ppm) = 31.3 (C$_i$), 40.2 (C$_h$), 42.3 (C$_b$), 106.9 (C$_f$), 114.8 (C$_c$), 145.1 (C$_e$), 148.3 (C$_g$).

IR (ATR, cm$^{-1}$): 3439 ($\nu$ N-H), 3348 ($\nu$ N-H), 3320 ($\nu$ N-H), 3061 ($\nu$ C-H$_{Ar}$), 2975 ($\nu_{as}$ C-H), 2920 ($\nu_s$ C-H), 1602 (C=C$_{Ar}$), 1476 ($\delta$ N-H), 1270 ($\nu$ C-N).

**Synthesis of tBuPhNH$_2$** : 1-(2,6-diamino-4-*tert*-butylbenzyl)-3-(3-(triethoxysilyl)propyl)urea

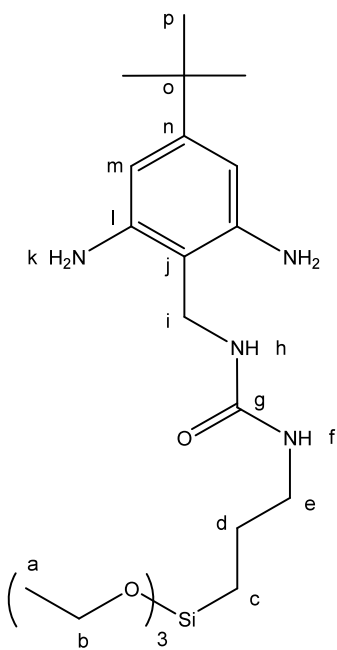

50 mg (0.26 mmol) of 2-(aminomethyl)-5-*tert*-butylbenzene-1,3-diamine were diluted in 10 mL of chloroform in a 50 mL round-bottom flask equipped with a cooler and a magnetic stirring bar. The vessel was sealed with a rubber septum, evacuated and backfilled with argon. 52 µL (0.21 mmol) of (3-isocyanatopropyl)triethoxysilane were then rapidly added with a syringe and the mixture was stirred at room temperature under argon for 15 h. The solvent was removed in vacuo and the oily residue poured into 20 mL of pure pentane (> 99 %). After sonication and decantation, the supernatant was removed with a syringe and then the residue was dissolved in 5 mL of dichloromethane and transferred into a Schlenk tube with a cannula. The solvent was finally evaporated. Analysis showed remaining traces of 2-(aminomethyl)-5-*tert*-butylbenzene-1,3-diamine which should not play any role during the grafting reaction. The obtained compound was a dark orange solid (55 mg, yield: 73 %). Caution: the product was sensitive to light and should be stored in the dark under nitrogen.

$^1$H NMR (200 MHz, DMSO, 300 K): $\delta$ (ppm) = 0.50 (t, 2H, $^3J$ = 8.1 Hz, H$_c$), 1.15 (t, 9H, $^3J$ = 7.3 Hz, H$_a$), 1.35 (s, 9H, H$_p$), 1.65 (p, 2H, $^3J$ = 8.1 Hz, H$_d$), 2.98 (dt, 2H, $^3J$ = 8.1, 5.7 Hz, H$_e$), 3.73 (q, 6H, $^3J$ = 7.3 Hz, H$_b$), 4.04 (d, 2H, $^3J$ = 5.8 Hz, H$_i$), 6.02 (s, 2H, H$_m$), 6.08 (t, 1H, $^3J$ = 5.7 Hz, H$_f$), 6.18 (s, 4H, H$_k$), 6.48 (t, 1H, $^3J$ = 5.8 Hz, H$_h$).

$^{13}$C{$^1$H} NMR (200 MHz, DMSO, 300 K): $\delta$ (ppm) = 7.7 (C$_c$), 18.4 (C$_a$), 24.8 (C$_d$), 33.0 (C$_p$), 39.5 (C$_o$), 43.0 (C$_i$), 44.1 (C$_e$), 59.5 (C$_b$), 107.8 (C$_m$), 116.2 (C$_j$), 143.1 (C$_l$), 149.2 (C$_n$), 157.6 (C$_g$).

IR (ATR, cm$^{-1}$): 3354 ($\nu$ N-H broad), 3060 ($\nu$ C-H$_{Ar}$), 2977 ($\nu_{as}$ C-H), 2926 ($\nu_s$ C-H), 1632 ($\nu$ C=O), 1563 (C=C$_{Ar}$), 1452 ($\delta$ N-H), 1249 ($\nu$ C-N), 1079 ($\nu_{as}$ Si-O-C), 958 ($\nu_s$ Si-O-C).